\documentclass[prb,a4paper,showpacs,twocolumn,superscriptaddress,longbibliography]{revtex4-2}
\usepackage[utf8]{inputenc}

\usepackage{graphicx}
\usepackage{fullpage}
\usepackage{amsmath}
\usepackage{amssymb}
\usepackage{bm}
\usepackage{hyperref}
\usepackage{cleveref}
\usepackage{svg}
\usepackage{braket}

\usepackage{inputenc}

\begin{document}
  
\title{Holonomic quantum manipulation in the Weyl Disk}
\author{Victor Boogers}
\author{Janis Erdmanis} \author{Yuli Nazarov}
\affiliation{Kavli Institute of Nanoscience, Delft University of Technology, 2628 CJ Delft, The Netherlands}
\begin{abstract}
It has been shown that a Weyl point in a superconducting nanostructure may give rise to a Weyl disk where two quantum states are almost degenerate in a 2D manifold in the parametric space. This opens up the possibility of a holonomic quantum manipulation: a transformation of the wave function upon adiabatic change of the parameters within the degenerate manifold. In this paper, we investigate in detail the opportunities for holonomic manipulation in Weyl disks.

We compute the connection at the manifold in quasiclassical approximation to show it is Abelian and can be used for a phase gate. To provide a closed example of quantum manipulation that includes a state preparation and read-out, we augment the holonomic gate with a change of parameters that brings the system out of the degenerate subspace.   For numerical illustrations, we use a finite value of quasiclassical parameter and exact quantum dynamics. We investigate the fidelity of an example gate for different execution times.

\end{abstract}

\maketitle
\thispagestyle{empty}

\section{Introduction}

A quantum computer promises unprecedented advantage over classical one for a set of challenging problems like protein folding\cite{Wecker2014} and prime number factorization\cite{Shor1999}. However, building it is a challenge since it is necessary to isolate a set of quantum states from the environment as well as to manipulate it within the same environment. A particular obstacle is a decoherence whereby the states pick up the fluctuations of the environment, which adds stochastic dynamical phases and may induce dissipation\cite{Bennett2000, Nielsen2009}.

An alternative way to meet the challenge is the quantum manipulation within degenerate subspaces that have gained considerable interest nowadays\cite{Zanardi1999, Carollo2016, Wu2005}. The resonant manipulation that can be  applied in the most systems with well-separated energy levels, does not work for degenerate subspaces. Instead, the manipulation is performed within a degenerate manifold: a set of parameters within which the states are degenerate and distinct. An adiabatic change of parameters in time along a trajectory within this manifold corresponds to a unitary operator in the degenerate subspace, that depends on the trajectory rather then on the way it is traversed.
Such mapping of a trajectory to a unitary operator for manifold is often characterized with a connection that sets how a wavefunction defined at a point is transported to other points in an infinitesimally small neighbourhood. 
An important property of a connection is the holonomy: a property that transporting a vector over different trajectories with the same starting point and destination results in different vectors. 
This is the manifestation of either curvature of the connection 
or a singularity that separates the trajectories into topological classes, so that the transportation results differ only if two trajectories belong to different classes.

 In quantum mechanics, the connection is characterized by a gauge potential\cite{Wilczek1984}, and the unitary operators correspond to the path integrals involving the potential. They are called holonomic transformations. The advantage of using these transformations for wave function manipulation is that in adiabatic limit the result is determined by the trajectory only, rather then by details of the time dependence of the parameter evolution along the trajectory. This provides the robustness against parametric noise and other decoherence sources.

In quantum mechanics, the connection of a manifold can be either Abelian or non-Abelian. The Abelian connection has been studied by Berry \cite{Berry1984} while Wilczek and Zee \cite{Wilczek1984} addressed the general non-Abelian case specific for degenerate manifolds. In general, a holonomic transformation is assigned to any curve in the Hilbert manifold. The tasformations have been studied in the context of geometric phase \cite{Aharonov1987}, and nonadiabatic geometric computation\cite{Sjoqvist2012, Leone2019, Abdumalikov2013, Economou2007, Li2017, Zhu2002, Nagata2018, Duan2001, Solinas2003}. For an Abelian connection, the unitary operators representing all trajectories with the same endpoints can be simultaneously diagonalized. Such diagonalization is not possible for a non-Abelian connection, and thus holonomy is irreducible. This in principle permits an implementation of a complete set of quantum gates to achieve universal holonomic quantum computation in a variety of systems\cite{Faoro2003}. However, the experimental realization of these schemes  appeared difficult due to long execution times required to get rid of non-adiabatic corrections\cite{Sjoqvist2012, Parodi2006, Toyoda2013}. 

A particular example of holonomic manipulation involves a two-dimensional system of  indistinguishable anyons. Changing the positions of the anyons provides a connection with a vanishing curvature\cite{Leinaas1977}.  The holonomic transformations are the same for all trajectories from the same homotopy class 
specified by number of windings around the anyons.
This sets the paradigm of topological quantum computation.\cite{Kitaev2003}  Within the paradigm, a quantum manipulation is implemented as a braiding, a move of anyons along the trajectories whereby all anyons return to starting points making loops around each other. The holonomic transformations in this case are robust against the fluctuations of the trajectory shapes, this promises to implement manipulations of  an exceptional fidelity\cite{Collins2006}. These opportunities have raised interest in anyonic excitations in solid-state systems such as Majorana superconductor-semiconductor nanowires\cite{Bommer2019} and specific fractional quantum Hall effect setups.\cite{Feldman2020, Nakamura2020}

Many quantum systems exhibit topologically protected energy-level crossings in three-dimensional parameter space that are commonly called Weyl points \cite{Herring1937}. In solid-state physics, the parameter space is a space of wave vectors confined to a Brillion zone of the crystal lattice. The Weyl points in solid-state band structures are a subject of active theoretical and experimental research\cite{Yan2017}. Another realization of Weyl points concerns a multi-terminal superconducting nanostructure, where the Weyl points appear as the crossings of mirror-symmetric Andreev bond states at zero energy\cite{Riwar2016} in the parametric space of three independent superconducting phases.

It has been shown recently that the interaction effects can cause a substantial modification of Weyl points.\cite{Erdmanis2018} A generic interaction model combines soft confinement and fluctuations in the parameter space. In the quasiclassical limit, a {\it Weyl disk} is formed in the vicinity of a point: two quantum states are (almost) degenerate in a finite two-dimensional region of the three-dimensional parameter space. 
The residual level splitting is exponentially small in the quasiclassical parameter. This makes the degeneracy physically achievable for a variety of systems. For instance, in a multi-terminal superconducting junction, the Weyl disk can be realized by placing large inductances between each superconductor terminal and the nanostructure. This makes the phase differences at the nanostructure softly constrained by the phase differences at the terminals. 

\begin{figure}[!b]
\begin{center}
	\includegraphics[width=\columnwidth]{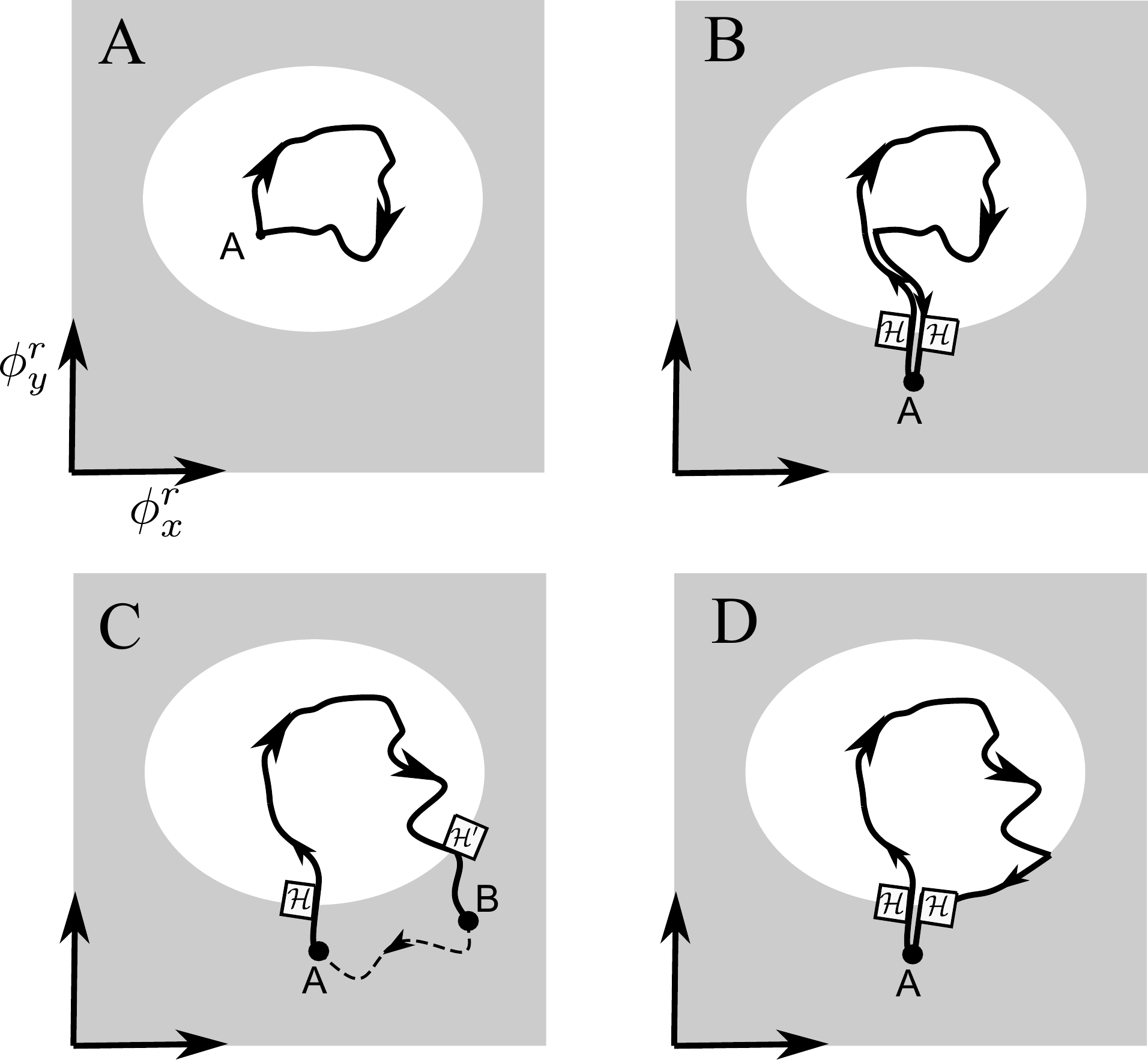} \\
	\caption{ A Weyl disk (shown in white) is a 2D elliptic region in the parametric space where two quantum states are degenerate. A. A purely holonomic transformation is achieved by adiabatic change of parameters along a closed trajectory. B. To achieve more functionality, we consider adiabatic passages beyond the degenerate maniold. A Hadamard gate describes the crossing of the disk boundary. C. The crossings of the boundary do not have to be in the same point. D. The gate from (C) is equivalent to the gate where the trajectory in the disk is closed along the disk boundary.}
	\label{fig:introduction}
\end{center}
\end{figure}

The degenerate manifold at Weyl disk may be used for holonomic quantum manipulation. We explore this opportunity in the present Article. Using the quasiclassical approximation, we compute the connection at the Weyl disk to show it is Abelian. A corresponding quantum gate (Fig. \ref{fig:introduction} a) is thus a phase gate in a proper basis. We show the relation of the phase shift and the Berry phase from the classic example of $1/2$ spin in magnetic field \cite{Berry1984}. To demonstrate richer opportunities for quantum manipulation, we augment the purely holonomic transformations by adiabatic passages to the exterior of the disk. (Fig. \ref{fig:introduction} b) We show that the crossing of the disk boundary corresponds to a Hadamard gate. 
With this, we provide a closed example of quantum manipulation that incudes an initialization in a superposition state,  holonomic manipulation, and subsequent readout.
 The crossings of the disk boundary can occur in different points as well (Fig. \ref{fig:introduction} c). The resulting gates are equivalent if a trajectory in the disk is closed along the boundary of the disk  
(cf. Fig. \ref{fig:introduction} c and d). We investigate the work of these quantum gates beyond adiabatic approximation with a full numerical simulation at a finite and moderate value of the quasiclassical parameter. We evaluate the gate fidelity as functio n of execution time and provide analysis of the dominant nonadiabatic corrections.

The paper is organized as follows. In Section \cref{sec:system} we provide a Hamiltonian description a multiterminal superconducting junction with a Weyl point and discuss the soft constraints that enable the Weyl disk regime. In Section \cref{sec:properties-weyl-disk} we consider the Weyl disk manifold in quasiclassical approximation and supplement it with a numerical example at a finite and moderate quasiclassical parameter. The Section \cref{sec:holon-transf} is separated into subsections where we (A) recall the concept of holonomic transformations, (B)compute the connection in the quasiclassical limit, (C) consider the adiabatic passages beyond the disk, (D) evaluate the connection beyond the quasiclassical limit. In  Section \cref{sec:even-odd-state} we analyze the deviations from adiabatic approximation, and present the results of the full quantum dynamics simulation evaluating the fidelity of the  swap gate as function of the gate execution time. We conclude in the Section \cref{sec:conclusions}.

\section{The system}\label{sec:system}

Weyl points in various physical systems have been a subject of an active research\cite{Obrien2019, Herring1937, Yan2017}. An important property of a Weyl point is its topological protection: the conservation of topological charge guarantees that small perturbations of the system just shift rather than destoy the point and associated conical singularity in energy spectrum.  

\begin{figure}[!b]
\begin{center}
	\includegraphics[width=\columnwidth]{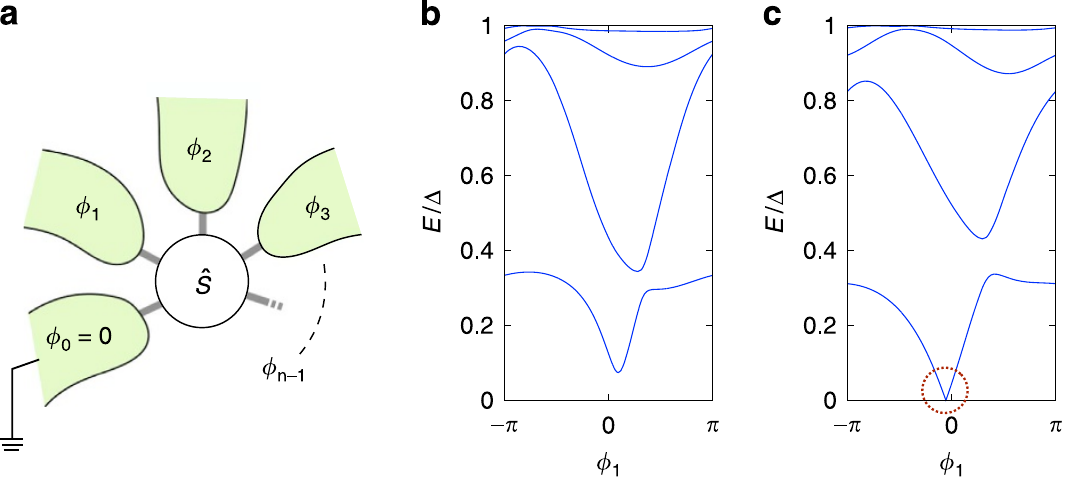} \\
	\caption{(Ref. \cite{Riwar2016}) \textit{Left}: Multiterminal superconducting junction with superconducting phases $\phi_n$. \textit{Middle}: The Andreev bound state spectrum as a function of one of the supeconductor terminal phases $\phi_1$ with the other  phases at general settings. \textit{Right}: The same spectrum for the choice of the other phases corresponding to a Weyl point.}
	\label{fig:MTJJ}
\end{center}
\end{figure}

Recently it was shown \cite{Riwar2016} that  multiterminal superconducting junctions with the leads of ordinary topologically trivial material can host Weyl points. In other words, the lowest in energy Andreev bound state (ABS) can be tuned to zero energy (Fig. \ref{fig:MTJJ}). The tuning parameters are the superconducting phases of the terminals. Owing to gauge invariance, only phase differences between terminals matter and thus at least four terminals are required to achieve Weyl points.

The Hamiltonian describing the conical spectrum in the  vicinity of a Weyl point is a $2\times 2$ matrix in the basis of two singlet degenerate ground states of the nanostructure \cite{Yokoyama2015},
\begin{equation}
H_{WP} =(\hbar/2e) \sum_{n=x,y,z} I_{n} \phi_n \hat\sigma_n.
\end{equation}
here, $\hat\sigma_n$ are the Pauli matrices in the space of the singlet states, $\phi_n$ are the superconducting phases counted from the positions, and  $I_n$ are the coefficients defining the energy slopes of the spectrum. The corresponding energy levels are 
$E = \pm (\hbar/2e)\sqrt{ \sum_n I^2_n \phi_n^2}$.

\begin{figure}[ht]
\begin{center}
  \includegraphics[width=\columnwidth]{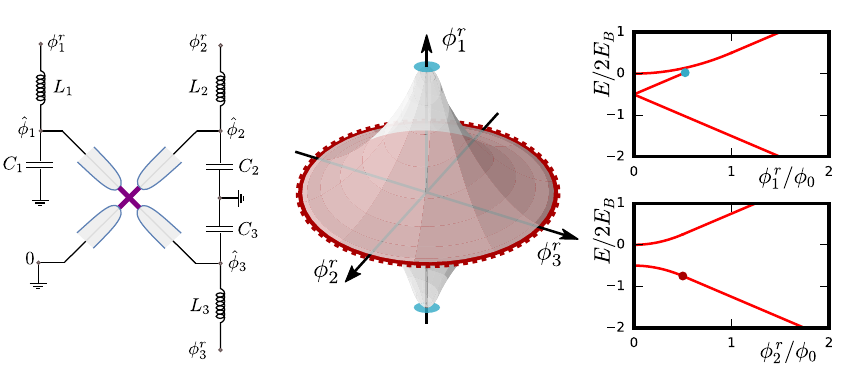} 
  \caption{(Ref. \cite{Erdmanis2018}). Formation of the Weyl disk. \textit{Left}: A four-terminal Josephson junction embedded in a linear circuit.  The linear circuit represents the finite capacitances and inductances of the superconducting leads and the surrounding electromagnetic environment. \textit{Middle} The 3D region where the  two minima of quasiclassical potential are present. The two minima are of the same energy at a 2D region shown in red: the Weyl disk. \textit{Right}: The energy spectrum along a line perpendicular to the disk (upper panel) and one in the disc plane (lower panel). In upper panel, the energies of two lowest minima split upon increasing the distance from the disk plane. In lower panel, two minima are degenerate and merge in one at the disk edge (red dot in the plot).  }
	\label{fig:WDSetup}
\end{center}
\end{figure}

In a realistic setup, the multiterminal superconducting junction is embedded in a linear circuit (Fig. \ref{fig:WDSetup}), and thus the phases determining the Weyl point  become dynamical variables rather than parameters, $\phi_n \to \hat{\phi}_n$, and can deviate from the external phases $\phi_n^r$ that play the role of parameters now. The linear circuit yet constrains softly the $\hat{\phi}_n$ to $\phi_n^r$. 

The constraint in general can be implemented as a quadratic addition to the energy. In particular, for our setup this is an inductive energy: each inductance $L_n$ adds a term $(\hbar/2e)^2(\hat\phi_n-\phi^r_n)^2/2L_n$ constraining the corresponding phase. The quantum  fluctuations of the phase around this point are determined by the corresponding capacitance of the system that produces charging energy $(2 e \hat {N}_n)^2/2C_n$, $\hat{N}_n$ being a variable canonically conjugated to $\hat{\phi}_n$. The full Hamiltonian of the system with a soft constraint reads:
\begin{eqnarray} \label{eq:Ham0}
	H(\vec{\phi^r}) &=& H_{WP} + \left(\frac{\hbar}{2e}\right)^2 \sum_{n}\frac{(\hat{\phi}_n-\phi_n^r)^2}{2L_n} \\
	&+& \sum_n \frac{(2 e \hat N_n)^2}{2C_n} \nonumber
\end{eqnarray}
where the middle term accounts for the soft constraint and the last term for the fluctuations. This is a minimum model of the embedding linear circuit, more complex models involve general frequency-dependent response functions of the circuit and do not change the qualitative conclusions.

The relevant scales in this Hamilonian can be understood  when considering a single-dimension version of it,
\begin{eqnarray} \label{eq:Ham1d}
	H(\phi^r) &=& (\hbar/2e) I \phi\hat\sigma_z + \left(\frac{\hbar}{2e}\right)^2 \frac{(\hat{\phi}-\phi_n^r)^2}{2L}  \\
	&+& \frac{(2 e \hat N)^2}{2C} \nonumber
\end{eqnarray}

The diagonalization of this Hamiltonian is trivial since the quasi-spin and $\phi$ separate and we have an oscillator centered at the positions that depend on the eigenvalue of spin $\sigma = \pm 1$, $\phi = \phi^r - \sigma \phi_0$, $\phi_0 \equiv 2 e IL/\hbar$. At $\phi^r=0$ these two positions correspond to degenerate minina separated by energy barrier $E_B=LI^2/2$. The energy spectrum is given by ($m$ being the number of quanta in the oscillator) 
\begin{equation}
E_{m,\sigma} = \hbar \omega (m+1/2) + \frac{\hbar}{2e} I \phi^r \sigma - E_B.
\end{equation}
where $\omega = \sqrt{LC}$ is the oscillator frequency. 

We assess the significance of quantum fluctuations by comparing the barrier height and the energy quantum in the oscillator. We introduce a quasiclassical parameter:
\begin{equation} \label{eq:QCP}
Q = \frac{E_B}{\hbar \omega} =\frac{1}{2} \left(\frac{LIe}{\hbar}\right)^2\frac{\hbar}{e^2Z},
\end{equation}
 $Z=\sqrt{L/C}$ being the characteristic impedance of the oscillator. Since $Ze^2/\hbar \simeq 10^{-2}$ for typical circuits, the parameter can be large even for relatively small inductances. If $Q\gg1$, the system is in the quasiclassical regime and the overlap of the states in two minima is exponentially small.  In the opposite limit $Q\ll 1$ the overlap is big and the effect of soft confinement can be treated perturbatively.
 
We see that for one-dimensional version of the Hamiltonian the energy levels retain conical singularity as far as its dependence on $\vec{\phi}^r$ is concerned. Generally, one expects this to hold for 3D case as well: the confinement would just renormalize the "velocities" $\partial E/\partial \phi_n$ defining the singularity. It has been discovered in \cite{Erdmanis2018} that there is an important exception from this general rule: for one of the directions --- we will call it an easy axis and take it for z direction --- the velocity remains finite while vanishing in the limit of large $Q$for two perpendicular direction. The easy direction is defined by the biggest energy barrier: $E_B^z \equiv L^z I_z/2 > E_B^x, E_B^y$. 

Thereby a Weyl point in the presence of confinement and in the quasiclassical limit becomes a Weyl disk (see Fig. \ref{fig:WDSetup}): two energy levels remain degenerate within an ellipse in the $\phi^r_x-\phi^r_y$ plane, the semiaxes being given by ($n=x,y$)
\begin{equation}\label{eq:7}
  A_n = \frac{4e}{\hbar I_n} (E^z_B - E^n_B).
\end{equation}
This is related to the existence of two potential minima in the vicinity of the Weyl point corresponding to two spin directions. In quasiclassical approximation, the corresponding wavefunctions do not overlap being localized near the minima.


It has been shown that the residual level splitting is exponentially small in the quasiclassical parameter. 
The approximate degeneracy in the Weyl disk is potentially interesting for quantum manipilation: that could allow a superposition of two degenerate states to evolve in time very slowly and enables holomorphic manupulations of this superposition by changing $\phi^r_x, \phi^r_y$ along a trajectory. In the next Section, we will discuss in detail the properties of the states at the Weyl disk. 

\section{Properties of the states at the Weyl disk}\label{sec:properties-weyl-disk}

In this Section, we complete and expand the analysis of the quantum states at the Weyl disk that was started in Ref. \cite{Erdmanis2018}. We present the numerical results and compare with the analytical ones in the quasiclassical limit.
This analysis is crucial for understanding the available holonomic transformations and other manipulations at the manifold.

\begin{figure}[h!]
  \begin{center}
    \includegraphics[width=\columnwidth]{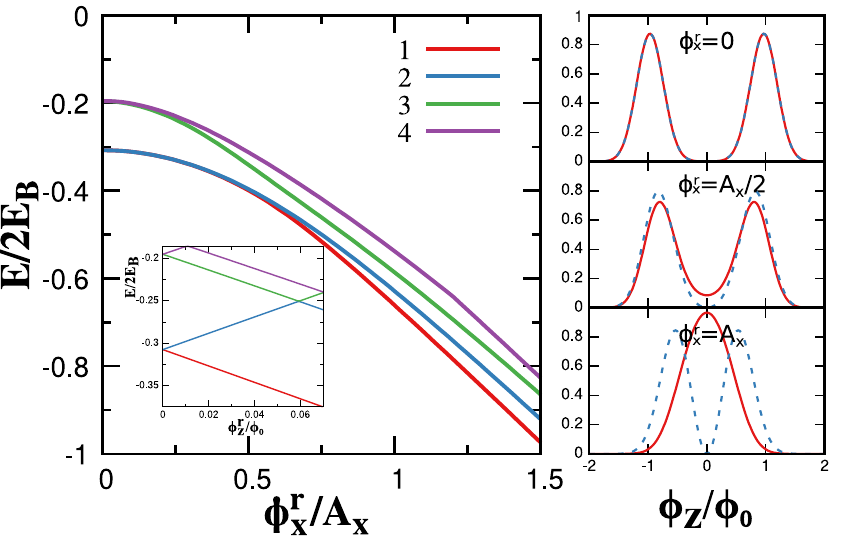}
    \caption{Numerical results for the energy spectrum and wave functions. For numerical illustrations in this Article, we choose a moderate value of the quasiclassical parameter $Q=E_B/\hbar \omega_z = 5$. ($E_B \equiv E_B^z$). Other parameters are $\hbar \omega_{x,y}/E_B = 1/3$, $E_B^{x,y}/ E_B = 1/3$. Left pane: Four lowest energy levels of the Hamiltonian \cref{eq:Ham0} in the disk plane $\phi^r_z=0$ versus $\phi^r_x$. The two lowest levels are degenerate deep in the disk. In the limit $Q \to \infty$ the degeneracy persists till the disk edge $(\phi^r_x)^2/A_x^2 + (\phi^r_y)^2/A_y^2 = 1$ with $A_{x,y}$ given by Eq. \cref{eq:7}. Since $Q=5$ is taken, the residual splitting at the disk edge is already comparable with $\omega_{x,y,z}$ which defines the energy distance to the higher levels. However, the splitting is not visible at $\phi^r_x < 0.5 A_x$.   
	Inset: 
	Four lowest energy levels versus $\phi^r_z$ at $\phi_{x,y}=0$. The splitting is lifted upon a shift $\phi^r_z$ in the easy direction and is proportional to the shift.
	Right pane: 
	The probability density of $\phi_z$, $p(\phi_z) = \sum_{\sigma}\int |\Psi_\sigma|^2 d\phi_x d \phi_y$ in the disk plane for even (solid line) and odd (dashed line) lowest energy states at several values of $\phi^r_x$.
	The probability deep in the disk separates in two almost non-overlapping peaks corresponding to two degenerate minima in the effective potential. 
	At the edge of the disk, the minima merge resulting in a single peak. 
    }
    \label{fig:E}
	\end{center}
\end{figure}

Let us first consider a numerical illustration (Fig. \ref{fig:E}). We choose a moderate value of the quasiclassical parameter $Q=E^z_B/\hbar \omega_z = 5$ in the Hamiltonian (\ref{eq:Ham0}). The barriers in $x,y$ directions are lower and equal, $E_B^{x,y}/ E_B = 1/3$.
The corresponding oscillator frequencies are also the same, $\hbar \omega^{x,y} = E_B^{x,y}$, this suggest the circular symmetry of the setup with respect to rotations about $z$. The effective $Q$ in this direction is thus $\simeq 1$. We will use this set of parameters for all numerical illustrations in the paper, since it proves the feasibility of holonomic manipulations at moderate values of the quasiclassical parameter.

In the left pane and inset of the Figure, we present the energies of the four lowest states. We will perform quantum manipulations in the basis of the two lowest states. As we will see later in dynamical simulations, the wave function from this basis mostly leaks to the 3rd and 4th state.

In the left pane, we plot the energies versus $\phi^r_{x}$ at $\phi^r_{z,y} =0$, that is, in the disk plane. The energies of the two lowest states are apparently degenerate up to $\phi^r_{x} \simeq 0.5 A_x$. Deep in quasiclassical limit, they remain degenerate up to $\phi^r_{x} =  A_x$. The moderate value of $Q$ results in the residual splitting that becomes comparable with the energy distance between the 3rd and the 2nd state at the edge of the disk. 

It is important for further consideration to note a symmetry of the Hamiltonian at this choice of the parameters: it is invariant with respect to $180^{\circ}$ rotation about $x$-axis. Owing two this, the splitting is diagonal in the basis of odd and even states with respect to the rotation. The 1st and the 4th state are even, while the 2nd and 3rd are odd. 

If we change the phase in easy direction, $\phi^r_z$, the degeneracy is immediately lifted (inset of the Figure), the splitting being proportional to $\phi^r_z$.
An instructive picture to comprehend this is that of a two-well potential depending on $\phi_z$. The energies of the distinct potential minima are aligned in the disk plane and are shifted by $\phi^r_z$ in opposite directions. The distance between the mimima reaches maximim at the center of the disk and decreases upon moving to the edge of the disk where two minima merge into one. In the quasiclassical limit, the wave function is localized at the minima. This is illustrated in the right pane of the Figure where we plot the probability density $p(\phi_z)$ for even and odd state. We observe two distinct peaks at the center, slightly overlapping peaks at $\phi^r_{x} \simeq 0.5 A_x$ and a single peak at the disk edge.


To characterize the double-well potential, we resort to quasiclassical approximation. If we neglect the quantum fluctuations completely, the quantum states are localized in superconductor phase space. The states can be decomposed as follows:
\begin{equation} \label{eq:AppFixedSpin}
  \ket{\Psi} = \ket{\vec{\phi}} \otimes \ket{S}
\end{equation}
where $\ket{\vec{\phi}}$ is a wave function with definite values of the superconducting phases and $\ket{S}$ is a 2-component wave function in the spin space. 
This decomposition allows us to determine $\ket{S}$ from the Hamiltonian (\ref{eq:Ham0}). Since we are looking for the states of the minimum energy, we set the spin to be antiparallel to the effective "magnetic field" at the point $\vec{\phi}$, 
\begin{equation} \label{eq:SpinCond1}
  \vec{W}\cdot \vec{\sigma} | S \rangle = -  | S \rangle
\end{equation}
$\vec{W}$ being the normalized vector in  the direction of the "field",
\begin{equation} \label{eq:defW}
W_n = \frac{I_n \phi_n}{\sqrt{\sum_n I_n^2 \hat \phi^2_n}}.
\end{equation}

\begin{figure}
          \includegraphics[width=\columnwidth]{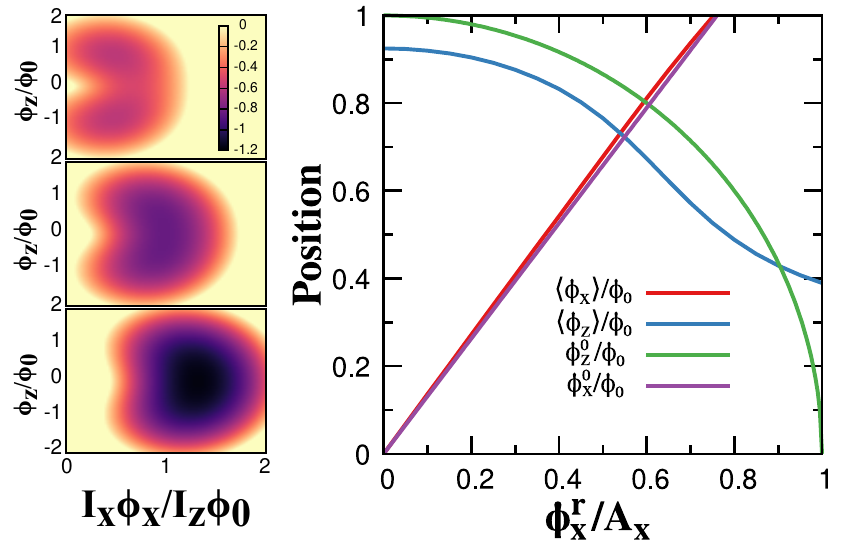}
          \caption{
The effective potential (Eq. \ref{eq:E_cl}) and comparison with the numerical results. Left pane: colour contours of the quasiclassical effective potential in the $\phi_z-\phi_x$ plane for the parameters at the Weyl disk:  $\phi^r_z,\phi^r_y=0$ and $\phi^r_x/A_x=0.5,1,1.5$ from the upper to lower plots. The minima of this double-well potential merge at the disk edge. Right pane: The locations of the potential minima $\phi^0_{x,z}$ given by Eq. \cref{eq:phi0_z}  and the numerical averages $\langle \phi_{x,z} \rangle$. To compute the averages, we take the wave functions of the two lowest states and arrange their superposition corresponding to the state localized in the upper minimum. The locations and averages are in reasonable agreement even for the moderate $Q$ chosen.}\label{fig:Potential}
\end{figure}

With this, we evaluate the effective potential for the state $\ket{\psi}$ aweraging the Hamiltonian over the state \cref{eq:SpinCond1} and neglecting the charging energy:
\begin{equation} \label{eq:E_cl}
  V(\vec{\phi}) = - \frac{\hbar}{2e} \sqrt{\sum_n I_n^2  \phi_n^2} + \left( \frac{\hbar}{2e} \right)^2 \sum_n \frac{(\phi_n - \phi_n^r)^2}{2 L_n}
\end{equation}

In Figure \ref{fig:Potential} (left) we plot the color contours of this effective potential in the $\phi_x-\phi_z$ plane  at three values of $\phi^r_x$ at  $\phi^r_z, \phi^r_y=0$. We see the minima moving towards each other upon increasing of $\phi^r_x$ and eventually merging at $\phi^r_x = A_x$. We expect the wave functions to be localized in the minima. To check for this, we compare the positions of the quasi-classical mimima with the numerical averages  $\langle \phi_{x,z} \rangle$ to find the reasonable correspondence even for the moderate $Q$. 

The positions of the minima are given by \cite{Erdmanis2018}:
\begin{align} 
  \langle \hat \phi_x \rangle = \frac{\phi_0 I_z}{A_x I_x} \phi^r_x && \langle \hat \phi_y \rangle = \frac{\phi_0 I_z}{A_y I_y} \phi^r_y \nonumber
\end{align}
\begin{equation} \label{eq:phi0_z}
  \langle \hat \phi_z \rangle = \pm \phi_0 \sqrt{1 - \rho^2}
\end{equation}
where $\phi_0 = (2e/\hbar) I_z L_z$, $\rho^2 = (\phi^r_x)^2/A_x^2 + (\phi^r_y)^2/A_y^2$ ($\rho^2=1$ corresponds to the disk edge). 

\begin{figure}[ht]
          \includegraphics[width=\columnwidth]{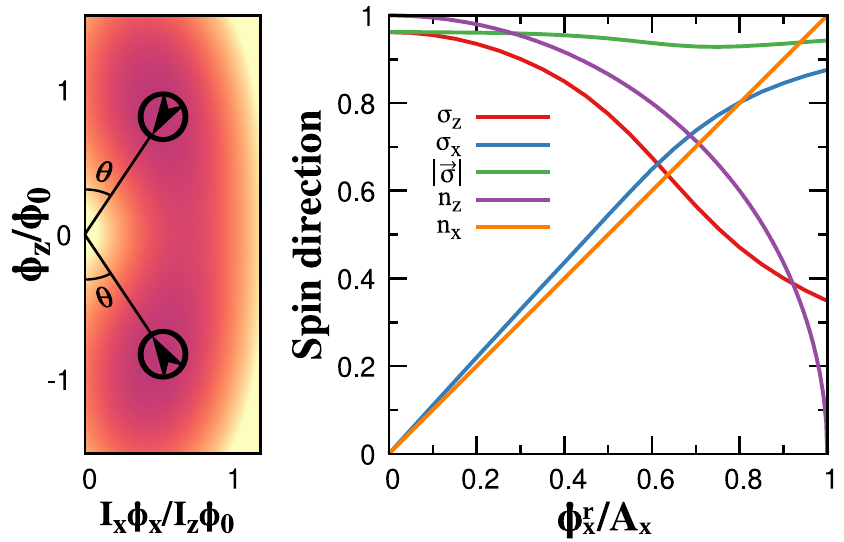}
          \caption{
The spin directions of the localized states compared with the numerical averages. Left pane: The countours of the even wave function. The spin directions in the potential minima are shown by the arrows. We rescaled the axes by $A_x, A_z$ so that the spin directions point towards the origin.  Right pane: the quasiclassical predictions for $\sigma_x,\sigma_z$ as compared with the numerical averages for the superposition of two lowest eigenfunctions corresponding to the localization in the upper minimum. we show the quasiclassical spin direction $\vec n$ with the spin direction obtained numerically. We normalize  $\langle \vec \sigma \rangle$ to $\langle \vec \sigma \rangle^2 =1$ to validate the decomposition given by Eq. \cref{eq:AppFixedSpin}. There is a good agreement even for the moderate $Q$ taken.}\label{fig:spinCond}
\end{figure}

The locations of the minima determine the spin of the localized states. With Eqs. \ref{eq:SpinCond1} and \ref{eq:phi0_z} we evaluate the angle $\theta$  it makes with the easy direction: $\cos \theta = \pm \langle \hat \phi_z \rangle/\phi_0$. The spins of two localized states in the center of the disk are aligned with $z$ and are antiparallel. At the disk we come closer to the disk boundary, the  the spins align with the disk plane and are parallel. (See Fig. \ref{fig:spinCond}.  The dependence of spin directions on $\vec{\phi}^r$  is essential for the holonomic transformations as discussed in the next Section.
To veryfy this numerically, we compute numerically the average values of $\sigma_x, \sigma_z$ for the superposition of the two lowest states that corresponds to the localization in  the upper minimum. The results are presented in the right pane of Fig. \ref{fig:spinCond}. We see a reasonable agreement with the quasicalssical results that is expectedly gets worse near the edge of the disk where the overlap of the states localized in different minima is significant.

\section{Quantum gates}\label{sec:holon-transf}

\subsection{Geometric phase and holomonic transformations}

Let us recall here the basics of geometric phase and holonomic transformations. We consider a  Hamiltonian $\hat{H}(\vec{x})$ 
that depends on a set of parameters $\vec{x}$. We change the parameters in time along a trajectory $\vec{x}(t)$, this results in the time-dependent Hamiltonian 
$\hat{H}(\vec{x})$. We introduce a local basis that diagonalizes  $\hat{H}(\vec{x})$ in each point of parameter space, 
\begin{equation}
E_n(\vec{x})|n\rangle =  \hat{H}(\vec{x}) |n\rangle
\end{equation}
The Schr\"{o}dinger equation in the local basis reads
\begin{equation}
i \hbar \dot{\psi}_n = E_n  - \hbar \dot{x}_i M^{i}_{nm} \psi_m 
\end{equation}
where the effective vector potential $\hat{M}^i$
\begin{equation}
\label{eq:connection}
M^{i}_{nm} = -i \langle n|\partial_i (|m \rangle) 
\end{equation}
represents the connection of the bases. 
With this, an in general, the unitary transformation of the wave function can be separated into two parts\cite{Aharonov1987}: the  dynamical phase arising from the first term representing the time-dependent energies and a geometric phase arising from the second term that depends solely on the trajectory in the parametric space. The geometric phase is an attractive phenonemon to use in quantum information processing. There are setups where the dynamical phase can be neglected beyond the adiababic approximation. Such examples have been actively studied in the field of nonadiabatic geometric quantum manipulation\cite{Sjoqvist2012, Leone2019, Abdumalikov2013, Economou2007, Li2017, Zhu2002, Nagata2018, Duan2001}. Alternatively, the trajectory in the parameter space can be passed adiabatically. The resulting geometric phase then reduces to a holonomic phase and is a basis of actively studied holonomic quantum manipulation\cite{Carollo2016, Wu2005}.

The most common example of adiabatic manipulation is the Berry phase \cite{Berry1984}. In this case, the energy levels are non-degenerate, and the adiabaticity implies that the frequencies associated with the parameter change are much smaller than the energy distances between the levels. One can neglect the non-diagonal elements of $\hat{M}^i$ so the Shr\"{o}dinger equation reduces to 
\begin{equation}
i \hbar \dot{\psi}_n = E_n(\vec{x}(t))\psi_n  - \hbar \dot{x}_i M^{i}{n} \psi_n 
\end{equation}
The dymanical phase separates from geometric phase. The latter depends on the trajectory only and is given by the line integral over the trajectory, $\vec{x}_i, \vec{x}_f$ being the initial and final point of the trajectory,
\begin{equation} \label{eq:geoPhase}
	\beta_n = \int_{\vec{x}_i}^{\vec{x}_f} \vec{M}_n \cdot d\vec{l}
\end{equation}
The vector potential is not gauge invariant and changes upon a gauge transformaton $|n\rangle \to U(\vec{x}) |n\rangle$, $|U|^2 =1$,
\begin{equation}
M^i_n \to M_n^i + U^*\partial_i U
\end{equation}
The guage-invariant Berry phase is defined for closed trajectories $\vec{x}_i = \vec{x}_f$ and, by virture of Stoke's theorem, equals to the surface intergal of the curl of $\vec{M}$ over the surface enclosed by the trajectory.

A fascinating extension of this concept petrains the case where several levels of the Hamiltonian are degenerate in a subspace of $\vec{x}$.
The adiabaticity implies that the frequencies of the change are much smaller than the energy distance between the degenerate and non-degenerate levels. 
The adiabatic motion along a trajectory results in a unitary transformation in the degenerate subspace $\hat{S}(\vec{x}_i, \vec{x}_f)$, 

\begin{equation} \label{eq:U-M}
\hat{S} = \mathcal{P}\exp\left(i\int_{\vec{x}_i}^{\vec{x}_f} \vec{\hat{M}}(\vec x) \cdot d \vec l\right). 
\end{equation}

$\mathcal{P}$ stands here for the ordering of $\hat{\vec{M}}$along the trajectory. The operator vector potential is not invariant with respect to the unitary transformations of the basis, $|n\rangle \to U(\vec{x}) |n\rangle$, $|U|^2 =1$,
\begin{equation}
M^i_n \to \hat{U}^\dagger M_n^i\hat{U} + \hat{U}^\dagger \partial_i \hat{U}.
\end{equation}
The gauge invariance is achieved for closed trajectories, and pertains the eigenvalues of $\hat{S}$.

The holonomic transformations can be Abelian and non-Abelian. They are Abelian if $\hat{M}^i(\vec{x})$ can be chosen to commute for all $\vec{x}$. As we will see soon, this will be the case under consideration. The non-Abelian connection permits  to realize a universal set of quantum gates \cite{Carollo2016, Wu2005} from holonomic transformations over different trajectories. However, the experimental implementation of them is difficult because of the nonadiabatic corrections\cite{Sjoqvist2012, Parodi2006, Toyoda2013}.


\subsection{Pure holonomic trasformation}
Let us apply these general considerations to the manifold of nearly degenerate wave functions at the Weyl disk. The first step is the parametrization of the basis in the degenerate subspace. We restrict ourselfs to the deep quasiclassical limit. The natural basis choice are that of wave functions localized either in upper ($\phi_z >0$) or lower  ($\phi_z >0$) minimum. As discussed, those can be decomposed into spin and orbital part,
\begin{equation}
|+\rangle = |S\rangle_+ |O\rangle_+;\; |+\rangle = |S\rangle_- |O\rangle_-
\end{equation}
Here, $|O\rangle_\pm$ are the normalized wave functions in $\vec{\phi}$ space located at the minima positions $(\langle \hat \phi_x \rangle , \langle \hat \phi_y \rangle, \pm\sqrt{1-\rho^2} )$ (see Eq. \ref{eq:phi0_z}). The $|S\rangle_\pm$ are spinors representing the spin antiparallel to the corresponding $\vec{w}_{\pm}$ (see Eq. \ref{eq:defW}), $\vec{w}_- = (w^x_+, w^y_+, - w^z_+)$. 

The best choice of the coordinates in the elliptic disk  corresponds to an unambiguous mapping of $(\phi^r_x, \phi^r_y)$ to the upper hermisphere of the vector $\vec{w}_+$. The two parametirzing angles $\theta$, $\alpha$, $0<\theta<\pi/2, -\pi<\alpha<\pi$ are determined from
\begin{equation}
\sin \theta = \rho; \; e^{i\alpha} = \left(\frac{\phi^r_x}{A_x} +i \frac{\phi^r_y}{A_y} \right) \rho^{-1};
\end{equation}
while $\vec{w}_+ = (\cos\alpha \sin \theta, \sin\alpha \sin \theta, \cos\theta)$.

It is essential to choose $|S\rangle_\pm$ to insure the continuity over the hemishere and the absence of singularity. This is achieved by setting 
\begin{equation}
|S\rangle_+ = \begin{bmatrix} - e^{-i\alpha} \sin\frac{\theta}{2}\\ \cos\frac{\theta}{2} \end{bmatrix}; \; 
|S\rangle_- = \begin{bmatrix} \cos\frac{\theta}{2}  \\ -e^{i\alpha} \sin\frac{\theta}{2} \end{bmatrix}.
\end{equation}

We compute the connection from Eq. \ref{eq:connection}. We may neglect the overlap
between $|O\rangle_+$ and $|O\rangle_-$ in the quasiclassical limit, so the connection is diagonal in this basis and therefore Abelian. Moreover, the derivatives of $|O\rangle_\pm$ with respect to $\theta,\alpha$ may be neglected as well. They give rise to the quantities proportional to the expectation values of the momentum and angular momentum for these states, those are zero since the states are localized. The connection is thus determined by the derivatives of $|S\rangle_+$
and reads:
\begin{equation}
\hat{M}^\alpha = -\hat{\tau}_z \sin^2\frac{\theta}{2}; \; \hat{M}^\theta =0
\end{equation}
Thereby we reduce the situation to the classic example of Berry phase for an electron spin in spin magnetic field of constant amplitude \cite{Berry1984}. 
Any holonomic transformation has a form of $\exp(-i\tau_z \beta)$. This is a phase gate, whereby the states $|  \pm\rangle $ acquire opposite phase shifts $\mp \beta$, $\beta$ being the Berry phase from the example. For any closed trajectory, $\beta$ is thus the half of a solid angle enclosed by the trajectory on the hemisphere,
\begin{equation}
\beta = \oint {\rm curl} \vec{M} dS = \frac{1}{2} \oint \sin \theta d \theta d\alpha.
\end{equation}
In original coordinates, the connection and the curl read as follows:
\begin{eqnarray}
M^{x}&=& \frac{\phi^r_y}{\rho^2 A_y A_x} (1-\sqrt{1-\rho^2})\\
M^y&=&  -\frac{\phi^r_x}{\rho^2 A_y A_x}  (1-\sqrt{1-\rho^2}) \\
{\rm curl} \vec{M} &=& \frac{1}{2 A_x A_y\sqrt{1-\rho^2}} 
\end{eqnarray}

We will consider the deviations due to finite $Q$ in the Subsection $D$. In the next Subsection, we will present quantum gates that enable measuring of the result of holonomic transformations. 

\subsection{Beyond the disk}

The initialization and measurement of a quantum state at the degenerate manifold is questionable if ever possible in principle. To check if holonomic transformations work as supposed, we need to extend the quantum manipilation schemes. A simple way to achieve this would be to depart from the disk in easy direction. This leads to energy splitting  $|\pm\rangle$ and enables the measurement in this basis. However, with this measurement one cannot characterise the work of the phase gate predicted, since it does not alter the probabilities to be in $|\pm\rangle$. Besides, the states almost do not overlap: this makes it difficult to arrange their superposition. We need to do something different.

We propose to augment the purely holonomic transformations in the disk with adiabatic passages in the same plane that go beyond the degenerate manifold. (Fig. \ref{fig:introduction}) This will bring us to the basis of the ground and first excited states that is continuous an unambiguous in the exterior of the disk. The adiabatic passages in the exterior change the phase difference between these basis states (mostly this is the effect of dynamical phase) not affecting the probabilities. Since the states are distinguishable (e.g. they correspond to different currents in the superconducting leads given by energy derivatives with respect to $\vec{\phi}$), these propabilities can be measured. Such measurements can be done with Andreev bound state spectroscopy\cite{Zazunov2003, Shafranjuk2002, Janvier2015}. The resonant quantum manipulation is also possible since the energies are spilt and the wave functions of the states overlap. 

Let us find how the wave function is transformed between the interior and exterior bases upon crossing the disk boundary in the point $\theta = \pi/2, \alpha = \alpha_0$. We consider a transformation $R(\alpha_0)$: $180^{\circ}$ rotation about the axis $(\cos\alpha_0, \sin\alpha_0, 0)$that is in the direction of the spin-orientation vectors $\vec{w}_{\pm}$ at this point. For a circular-symmetric disk $L_x=L_y, I_x = I_y, C_x = C_y$ this is a true symmetry transformation of the Hamiltonian. For an anisotropic disk, this symmetry holds approximately in quasiclassical limit where the wave functions are concentrated near a point in $\vec{\phi}$ space. 

The transformation should be diagonal in exterior basis. The ground and excited state are respectively even and odd upon $R(\alpha_0)$. As to the interior states, let us note that $|0\rangle_{\pm} = R(\alpha_0) |0\rangle_{\mp}$, so that 
\begin{equation}
R(\alpha_0)|\pm\rangle = e^{\mp i \alpha_0} |\mp\rangle 
\end{equation}
With this, we find that the wave functions in exterior and interior bases are related by a generalized Hadamard gate
\begin{equation}
\mathcal{H}(\alpha_0) \equiv \frac{1}{\sqrt{2}} \begin{bmatrix} 1 & e^{i\alpha_0} \\ e^{-i\alpha_0} & -1 \end{bmatrix}
\end{equation}
Since $\mathcal{H}^2=1$, the same matrix relates the bases upon the reverse passage.

Let us consider the quantum gate given in Fig. \ref{fig:introduction} b. We initialize the wave function in a point $A$ beyond the disk: we can wait for the relaxation that brings the system to the ground state. After this, we can bring it to a superposition of the ground and excited state with a pulse of an oscillating $\vec{\phi}$ with the frequency matching the level splitting in this point. The adiabatic trajectory enters the disk, makes a loop there for a holonimic manipulation, and returns to the same point. The resulting quantum gate reads
\begin{equation}
\hat{S} = \mathcal{H}(\alpha_0) e^{i\tau_z \beta} \mathcal{H}(\alpha_0) 
\end{equation}
$\beta$ being the Berry phase accumulated on the loop. This does not include the phase shifts in the exterior basis that do not change the probability. If we start in the ground/excited state, we end up in the excited/ground state with the probability 
\begin{equation}
\mathcal{T} = \sin^2 \beta
\end{equation}
Measuring these probabilities thus permits the characterization of the holonomic transformation. The answer for the probability, as expected, does not depend on the entrance point $\alpha_0$. 

To measure the wave function in the exterior basis, one does not have to return to the initial point  (Fig. \ref{fig:introduction} c): the measurement can be performed upon leaving the disk at some other point $(\pi/2, \alpha_1)$. The resulting quantum gate upon the phase factors in exterior basis is given by 
\begin{equation}
\hat{S} = \mathcal{H}(\alpha_0) e^{-i\tau_z \beta} \mathcal{H}(\alpha_1) 
\end{equation}
so the probability ${\cal T} = \sin^2 (\beta^* +(\alpha_1 -\alpha_0)/2)$, $\beta^*$ being the Berry phase acquired upon the part of the trajectory that connects the entrance and exit points. One may be surprised with the fact that $\beta^*$ in principle is not guage invariant quantity. This is resolved if we note that in our gauge $(\alpha_1 -\alpha_0)/2$ is the Berry phase acquired upon a passage along the disk boundary from $\alpha_1$ to $\alpha_0$. So the gate in Fig. \ref{fig:introduction} c is equivalent to that in Fig. \ref{fig:introduction} d 
 where the trajectory in the disk is closed.
This restores the guage-invariant expression 
${\cal T} = \sin^2 \beta$, $\beta$ being the Berry phase accumulated along the closed trajectory.

More sophisticated gates can be arranged by entering and leaving the disk repeatedly along an adiabatic passage. They are composed of Hadamard gates, holomonic phase shifts in interior basis and dynamical phase shifts in exterior basis. 

\subsection{Connection beyond the quasiclassical limit}
The simple expression for holonimic transformation obtained above is valid in the deep quasiclassical limit only and relies on the localization of the wave functions. One may wonder how accurate it is at finite values of $Q$. At first site, this problem is superfluous since finite values of $Q$ give rise to the splitting of degenerate values in the disk, this formally invalidates the holonomic transformation. However, the splitting is exponentially small and may be neglected when the deviations from the deep quasiclassical limit are noticeable. 

To investigate and illustrate this, in this Subsection we compute numerically the connection at finite $Q$. We restrict ourselves to a simple particular case when this computation is straightforward: we concentrate on the circular trajectories at a circular-symmetric disk.   

The circular symmetry of the disk implies $I_{x,y}=I_r$, $L_{x,y}=L_r$, $C_{x,y}=C_r$. 
Let us concentrate on a family of the Hamilitonians 
$\hat{H}(\alpha)$ along a circular trajectory $\phi^r_x = r \cos\alpha, \phi^r_y = r \sin\alpha$, $r$ being the radius. 
This family is obtained by rotations about $z$ axis by $\alpha$, 
\begin{equation} \label{eq:HamRot2}
	\hat{H}(\alpha) = R^{-1}_z(\alpha) \hat{H} R_z(\alpha);
\end{equation}
where 
\begin{equation}
R_z(\alpha) = \exp( - i \alpha \hat{J}_z)
\end{equation}
and $\hat{J}_z$ is the angular momentum operator. 
$R_z^{\alpha}$ represent rotation around easy direction with an angle $\alpha$. The rotation thus generates the family of the bases
diagonalizing $\hat{H}(\alpha)$:
\begin{equation}
\label{eq:thebasis}
|n\rangle_\alpha = \exp( - i \alpha \hat{J}_z)|n\rangle_\alpha
\end{equation}
The connection $\hat{M}^\alpha$ is thus determined through the matrix elements of $\hat{J}_z$ (see Eq. \ref{eq:connection}), is constant over the trajectory. One needs to diagonalize the Hamiltonian only once per trajectory to compute the connection.

\begin{figure}[ht]
		\includegraphics[width=\columnwidth]{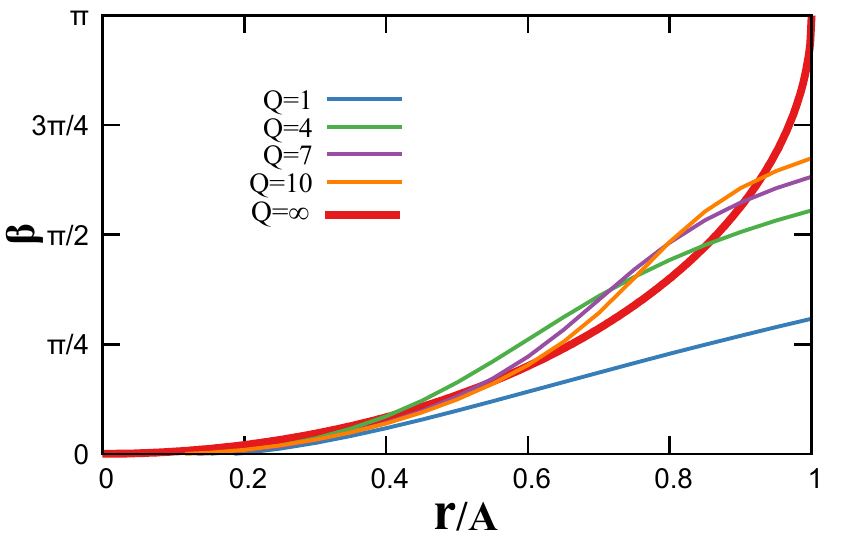}
		\caption{
                  Holonomic phase at various finite values of $Q$ for a circular trajectory of radius $r$ at the circular-symmetric disk. We compare it with the quasiclassical result (Eq.\ref{eq:betaquasi}) in the limit $Q \to \infty$. 
				  }
				  \label{fig:CirclesCapacities}
\end{figure}

We project $\hat{J}_z$ on the two lowest eigenstates $|e\rangle$,$|g\rangle$,  of the Hamiltonian, 
\begin{equation}
M^\alpha_{ab} = \langle a| \hat{J}_z | b \rangle
\end{equation}
$a,b = g,e$.The half-difference of the eigenvalues of this matrix defines the holomic transformation phase accumulated over the circular trajectory $\beta = \pi (M_+-M_)$.

One needs to take into account that the einvalues of $\hat{J}_z$ are half-integer and the basis given by Eq. \ref{eq:thebasis} is discontinious. Owing to this, a formal calculation would give $\beta = \pi$ even at $r \to 0$ where no change of the Hamiltonian takes place. So one has to subtract $\pi$ to make sure $\beta \to 0$ at $r \to 0$.

The results are plotted in Fig. \ref{fig:CirclesCapacities}. In the deep quasiclassical limit, the angular momentum operator can be replaced with $\sigma_z/2$ and the holonomic phase is given by
\begin{equation}
\label{eq:betaquasi}
\beta = \pi \frac{1-\langle \hat{\sigma}_z \rangle}{2} = \pi \left(1- \sqrt{1-(r/A)^2}\right),
\end{equation}
$A$ being the disk radius. We observe significant corrections to the quasiclassical limit at finite $Q$, those become bigger at the disk boundary and at the smaller $Q$. However, the overall dependence of $\beta(r)$ is preserved even at $Q=4$.

\section{Quantum dynamic}\label{sec:even-odd-state}

In this Section, we discuss the deviations from the ideal results of the execution of  the quantum gates described. The deviations come from the residual level splitting in the disk and from the non-adiabatic excitations to higher levels in the course of the execution at finite time. We illustrate these sources with the numerical examples of the quantum dynamics of the full Hamiltonian (\ref{eq:Ham0}). The parameters of the Hamiltonian used for illustrations are the same as in the previous Sections corresponding to the moderate quasiclassical parameter $Q$ and circular symmetry. We show that the gates work well even in this case.
\begin{figure}[!b]
\begin{center}
	\includegraphics[width=\columnwidth]{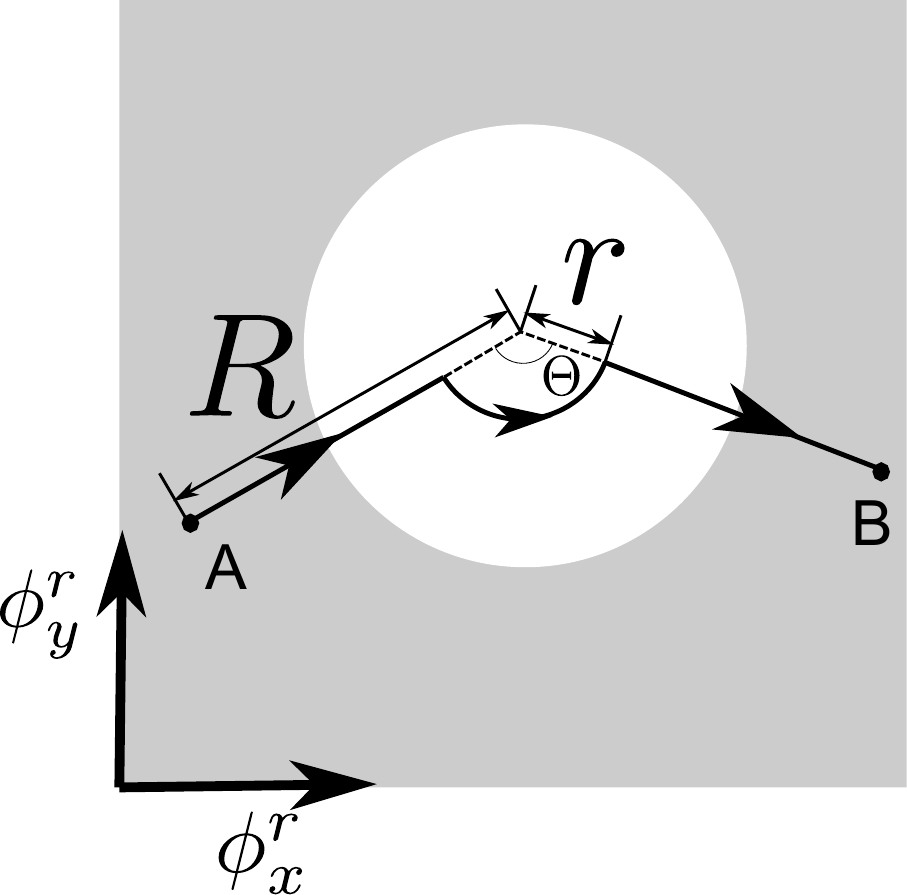} \\
	\caption{The concrete example trajectories investigated in Section \ref{sec:even-odd-state}.}
	\label{fig:passages}
\end{center}
\end{figure}


We concentrate on the gate of the type given in Fig. \ref{fig:introduction}c. We chose a simple family of the trajectories (Fig. \ref{fig:passages}) consisting of two straight passages ("arms")  in radial direction and an arc around the origin. The initial and final points $A,B$ are beyond the disk at the same distance $R>A$ from the origin, $A$ being the disk radius. The radius and the angle of the arc are $r$, $\Theta$ respectively. We expect the holonomic phase $\beta$ to accumulate upon the passage:
\begin{equation}
\beta = \frac{1}{2} \Theta \sqrt{1- \left(\frac{r}{A}\right)^2}.
\end{equation}

The time dependence of $\vec{\phi}^r$ at the trajectory can be defined in terms of the angular velocity $\Omega$ at the arc part and the linear velocity $v \equiv A \dot{\rho}$ The overall execution time of the gate is thus $T=\Theta/\Omega + 2(R-r)/v$. In a realistic manipulation, it is easy to make these velocities time-dependent preserving the total execution time, this might be a possibility to reduce the non-adiabatic corrections and thus improve the gate performance. However, the qualitative analysis made and our attempts of such optimization did not show any substantial improvement. The optimal time dependence of the velocities is close to constant. 

We consider the deviations at the arc part and at the arms separately, and conclude by combining both in an example of the gate fidelity versus the execution time $T$. 

The Schr\"{o}dinger equation at the arc part is best expressed in the local basis (\ref{eq:thebasis}),

\begin{equation}
i \hbar \dot{\psi}_n = E_n\psi_n  - \hbar \Omega(t) J^z_{nm} \psi_m 
\end{equation}

the effective Hamiltonian not depending on time if $\Omega(t)={\rm const}$. The initial condition corresponds to the wave function localized in the two lowest levels, $\psi_{1,2} \ne 0$, and the equation needs to be solved at the time interval $(0,\Theta/\Omega)$.

The first source of the deviations is the residual level splitting $E_2 -E_1$. The proper work of the holonomic gate requires this splitting to be smaller than the second term $\propto \Omega$, that is, $ |E_2 -E_1| \ll \hbar \Omega$. If this condition is fulfilled, the deviations in probabilities are proportional to 
$((E_2 -E_1)/\hbar\Omega)^2$. 

The second source are the non-adiabatic corrections corresponding to the excitations to higher levels $n>2$. The probabilities of the excitations from the states $|1\rangle, |2\rangle$ can be estimated as 
\begin{align}
  P_{1,2 \to n} \approx |\langle n |\hat{J}_z |1,2\rangle|^2 \left(\frac{\hbar \Omega}{E_n - E_{1,2}}\right)^2 
\end{align}
The small probabilities thus require $\hbar \Omega \ll |E_n - E_{1,2}| \simeq \hbar \omega_{x,y,z}$. The execution time thus should be 
\begin{figure}[ht]
  \centering

  \includegraphics[width=\columnwidth]{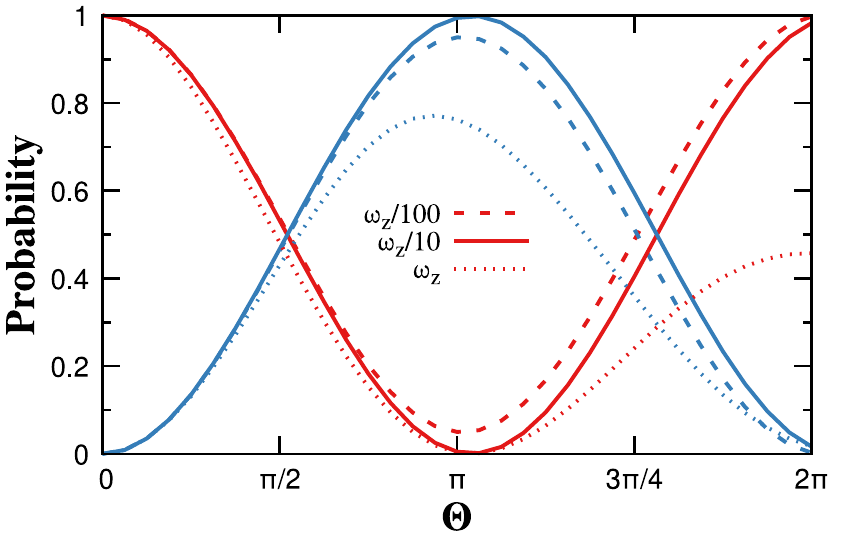}
  \caption{
    The deviations coming from the arc part. We assume the ideal work at the arms of the trajectory. The initial state is an excited state $|2\rangle$ in the exterior basis. We plot the probabilities $P_{1,2}$ (blue and red) versus $\Theta$ at three different angular velocities $\Omega = 10^{-3}, 10^{-2},10^{-1} \omega_z$ (dashed, solid and dotted lines). At the intermediate $\Omega$, the results coincide with those for the ideal gate upon numerical accuracy. At the smallest $\Omega$, the deviations are due to the residual level splitting in the disk interior. At the highest $\Omega$, the deviations are due to non-adiabatic excitiation to the higher levels.
	 }
	\label{fig:arc}
\end{figure}

We illustrate this with the quantum dynamics simulation at the arc part of the trajectory. (Fig. ref{fig:arc}). We simulate the work of the gate assuming its ideal execution at the arms. The initial state corresponds to the excited state in the exterior basis at the point $A$. We compute the probabilities $P_{1,2}$ to end up in the ground/excited state at the point $B$. For an ideal gate, those are given by $P_{1,2} = \sin^2\beta, \cos^2\beta$. 

We plot the probabilities versus $\Theta$ for three different angular velocities $\Omega = 10^{-2},10^{-1},10^{0} \omega_z$. 
We choose $r =0.25$ where the residual splitting $E_{2}-E_{1} = 0.002 \omega_z$. At the smallest $\Omega$, the deviation owing to the residual splitting is noticeable. There is no excitation to the higher levels, so that $P_1+P_2=1$. At the intermediate $\Omega$, the results follow those of the ideal gate with the numerical accuracy. The probabilities achieve minimum/maximum at $\Theta \approx \pi 1.03$ corresponding to $\beta=\pi/2$. At the highest $\Omega$, the non-adiabatic correction become noticeable. The probability progressively leaks to higher levels, with only a half of it remaining in the computational subspace for the longest gate $\Theta=2\pi$. Despite the significant deviations, the curves at the lowest and at the highest angular velocity still follow the oscillatory pattern. This demonstrates that the gate works in a wide range of the angular velocities.


Let us analyse the deviations coming from the radial arms of the trajectory. The residual splitting in this case only modifies the phase factor in the exterior basis and can be disregarded. To quantify the non-adiabatic corrections, we represent the Schr\"{o}dinger equation in the instantaneous basis $H(t) |n(t)\rangle = E(t) |n(t)\rangle$ and compute the probabilities to be in the excited states $n>2$ in the second order perturbation theory in terms of the non-diagonal elements of $ {\partial H}{\partial t}$. 

Owing to circular symmetry, the results do not depend on the direction of the arm. We can compute the amplitudes for the motion in $x$ direction with the velocity depending on the actual value of $\phi^r_x$, $\dot \phi^r_x(\phi^r_x)$ The amplitude in the excited state $n$ accumulated in the course of motion from $r$ to $R$ then reads:
\begin{multline}
  \psi_n(R) = - \frac{\hbar I_x}{2e} \int_r^R ds_2 \frac{|\langle n(s_2)| \sigma_x|a(s_2) \rangle|}{E_n(s_2) - E_a(s_2)}  \\
  \times \exp \left[ \frac{i}{\hbar} \int_r^{s_2} \frac{1}{\dot \phi^r_x(s_1)}(E_n(s_1) - E_a(s_1)) ds_1 \right]
\end{multline}
where $|a\rangle$ is a state from the computational subspace, either ground one or excited one.

\begin{figure*}[ht]
  \centering
  \includegraphics[width=1.8\columnwidth]{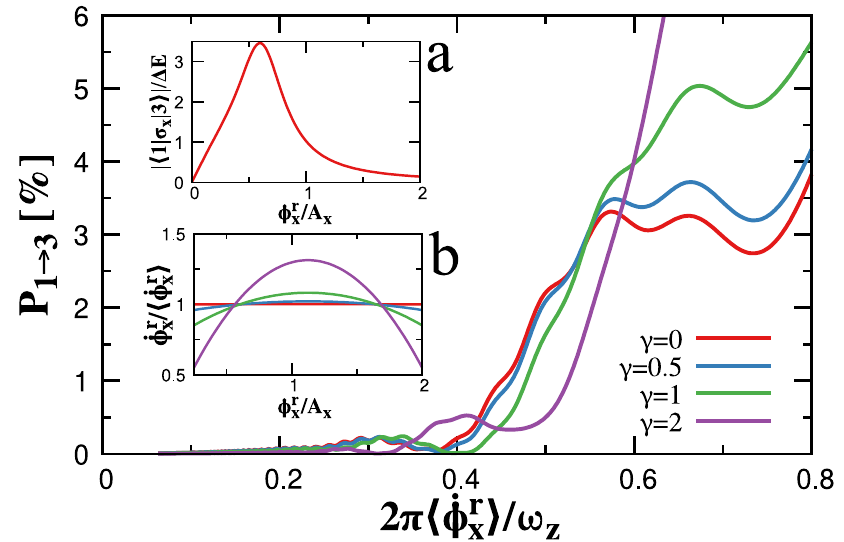}
  \caption{ The deviations coming from the radial part of the trajectory. The probability of the dominant excitation $P_{1\to3}$. For this examle, the trajectory starts at $r=0.25 A$ and ends at $R = 2 A$.
  a. The intensity of the matrix element energizing the transition versus the distance from the origin $\phi_x^r$.
  b. The velocity profiles corresponding to Eq. \ref{eq:profiles}.
  The curves correspond to various $\gamma$ shown in the labels.
  Main figure: we plot the probability for various velocity profiles ($\gamma$ is shown in the curve labels) as function of the average velocity. The probabilities are lower than $0.005$ for lower velocities $< 0.06 A \omega_z$ and increase rapidly at higher velocities. The slowest growth corresponds to the constant velocity profile $\gamma =0$.  
  }
  \label{fig:arms}
\end{figure*}

To evaluate this integral numerically, we use instantaneous eigenstates obtained by diagonalization of the Hamiltonian at various $\phi^{r}_{x}$. We summarize the results in Fig. \ref{fig:arms} where we present the probability $1 \to 3$   versus the average velocity at the arm. The non-diagonal matrix element energizing the transition is plotted versus $\phi^r_x$ in Fig. \ref{fig:arms} a.  We try a family  of $\phi^{r}_{x}$ dependencies of this velocity parametrized by $\gamma$:
\begin{equation}
\label{eq:profiles}
\frac{\dot \phi^r_x(\phi^r_x)}
{\langle  \dot \phi^r_x \rangle} = \frac{\gamma}{2 \tanh (\gamma/2)} (1 - \rho^2 \tanh^2(\gamma/2))
\end{equation}
where $\rho= (2\phi^r_x - R - r)/(R-r)$. These velocity profiles are plotted in Fig. \ref{fig:arms} b for several values of $\gamma$. 
In the main Figure, we plot the excitation probability versus the average velocity for several $\gamma$. The overall dependence is qualitatively consistent with exponential suppression of the transitions at low velocities, $\ln P \simeq v^{-1}$. The detailed dependence is not smooth, the probabilities oscillate showing interference due to finite length of the arm. At all velocity profiles checked, the probabilities are $< 5\cdot 10^3$ for  $\dot\phi^r_x \approx A \omega_z /4 \pi$ and grow rapidly at higher velocities. We find that the time-independent velocity profile $\gamma =0$ eventually provides smaller probabilities and is thus advantageous for the gate design.

\begin{figure}[ht]
  \centering
  \includegraphics[width=\columnwidth]{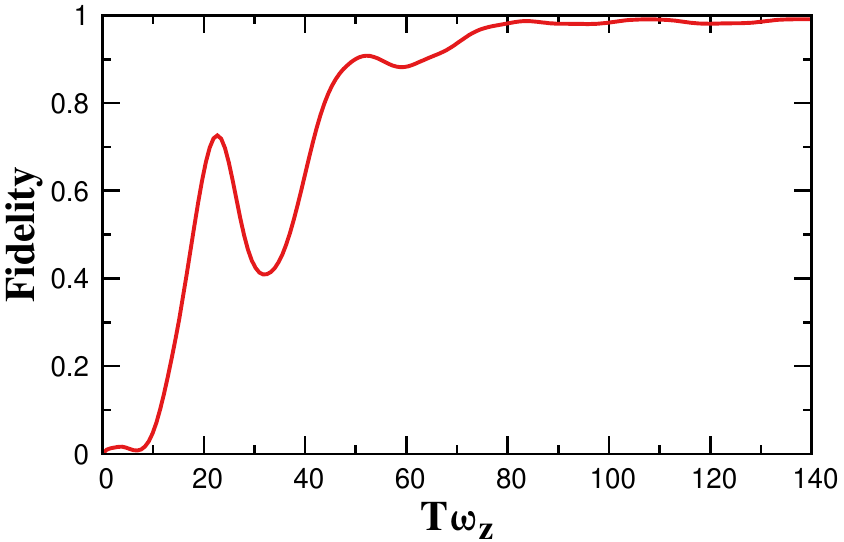}
  \caption{
    The fidelity of the holonomic swap gate as function of the execution time $T$. The gate corresponts to the trajectory with $r=0.25 A$, $R= 2A$, $\Theta = \pi$ and is executed with constant velocity along the trajectory.}
	\label{fig:swapgate}
\end{figure}

We illustrate the deviations on all parts of the trajectory by computing the fidelity of the swap gate versus the execution time. (Fig. \ref{fig:swapgate}). A swap gate implements the transformation $|1\rangle \to |2\rangle; |2\rangle \to |1\rangle$ in the exterior basis, that is, $\cal{T} = 1$. The actual example differs a bit from the swap gate: the trajectory parameters are $r =0.25 A$, $ R= 2 A$, $\Theta = \pi$, this corresponds to $\beta  
 \approx 1.5209$ that differs slightly from the swap gate value of $\pi/2$. However, ${\cal T} \approx 0.9975$ for this value of $\beta$ so the difference with the ideal swap gate is negligible.
The velocity in $\phi^r_x - \phi^r_y$ is constant along all parts of the trajectory. 
The fidelity is almost zero at $T\omega_z <10$, increases non-monotonically till $T\omega_z < 80$ and is close to ideal value at bigger $T$. We thus expect the gate to work good at $T\omega_z \simeq 100$ and longer. This looks parametrically bigger than a naive ad-hoc estimation $T\omega_z \simeq 1$. However, we have to take into account that the typical energy differences along the path are $\simeq 0.2 \omega_z$ and the better estimation for time is a period rather than the inverse frequency. So that  $T\omega_z \simeq 100$ corresponds to $3-4$ typical oscillation periods. Eventually, this corresponds to the number of peaks seen in the time dependence of the fidelity. 
 
In the figure, we show a full numerical simulation of the swap gate for different runtimes. As expected from small runtimes, the execution is diabatic and thus all state probabilities after execution are close to zero. On the other hand, at long runtime limit, we see that the gate is relatively stable a desirable property for holonomic computation. 

\section{Conclusions}\label{sec:conclusions}
In conclusion, we have investigated holonomic manipulations that can be performed utilizing approximate degeneracy at the Weyl disk: a rather counter-intuitive example of 2D finite degenerate manifold in 3D parameter space. The Weyl disks can be realized by soft confinement of parameters in the superconducting nanostructures hosting the Weyl points in the spectrum of Andreev bound states, which is the example considered here. The resulting Hamiltonian is however generic and can be realized in many quantum systems with the Weyl like crossings in the spectrum of discrete energy states, so our results are of general nature.

We have computed the connection in the Weyl disk manifold in quasi-classical limit and found it Abelian: it realizes a phase gate, the phase difference being related to the Berry phase in its classic example. This may seem a rather discouraging result. However, we propose to augment the purely holonomic transformations with the adiabatic passages beyond the degenerate manifold. With this, we can propose the realization of more sophisticated gates and practical measurement of the results of the holonomic transformations.

We did quantum dynamic simulation of the gates proposed for realistic Hamiltonians and find they can work properly at rather short execution times. We did not consider decoherence in this respect: the point is that our exemplary setup provides no projection against the fluctuations of the parameter along the easy axis that can cause the decoherence in the degenerate 2D subspace.

A natural continuation of this research line would include the consideration of several Weyl points brought into soft confinement and interaction with each other. They would give rise to degenerate manifolds of higher dimensions with richer holonomic transformations and perhaps provide the protection against decoherence.

To support open science and open software initiatives and to comply with institutional policies, we have published all relevant code and instructions for running it on the Zenodo repository\cite{HolonomicCode}. 

\acknowledgments

This research was supported by the European
Research Council (ERC) under the European Union's
Horizon 2020 research and innovation programme (grant
agreement No. 694272).

\bibliography{bibliography}

\end{document}